\documentclass[12pt]{article}
\usepackage[T1]{fontenc}
\usepackage[utf8]{inputenc}
\usepackage{amsmath}
\usepackage{amssymb}
\usepackage{amsthm}
\usepackage{graphicx}
\usepackage{float}
\usepackage{booktabs}
\usepackage{geometry}
\usepackage{hyperref}
\usepackage{url}
\usepackage{enumitem}
\usepackage{microtype}
\usepackage{array}
\usepackage{tabularx}
\usepackage{changepage}
\usepackage{caption}
\usepackage{mathpazo}
\newcommand{\eq}[2]{\begin{align}\label{#1}#2\end{align}}

\newcommand{\nn}{\nonumber}

\newcommand{\pa}{\partial}

\newcommand{\ep}{\epsilon}

\newcommand{\ga}{\gamma}\newcommand{\Ga}{\Gamma}
\newcommand{\la}{\lambda}
\newcommand{\om}{\omega}

\newcommand{\Eq}[1]{\eqref{#1}}
\begin{document}
	\title{Reconsidering the Vacuum Energy of a Massive Scalar Field in a Spherical Cavity}
	\author{Michael Bordag$^{1,2,}$* and Irina G. Pirozhenko$^{2}$}
	\date{}
	\maketitle
	\begin{center}
		\small
		$^{1}$ Institute for Theoretical Physics, Leipzig University, 04109 Leipzig, Germany;\\
		$^{2}$ Bogoljubov Laboratory of Theoretical Physics, Joint Institute for Nuclear Research, 141980 Dubna, Russia\\
		pirozhen@theor.jinr.ru\\[2pt]
		* Correspondence: bordag@mail.ru\\\date{Oct.18, 2026}
	\end{center}
	\begin{abstract}
		We reconsider the vacuum (Casimir) energy of a massive scalar field obeying Dirichlet boundary conditions inside a spherical cavity. Using zeta-function regularization, we carefully recalculate the analytic part of the zeta function via the uniform asymptotic expansion of the modified Bessel functions \cite{bord97-56-4896}. We correct a sign error present in an earlier calculation that affected the large-mass behavior. The renormalized vacuum energy is shown to vanish as $m\to\infty$, as required on physical grounds. Numerical results reveal that the energy changes sign several times as a function of the cavity radius.
	\end{abstract}
	\section{Introduction}\label{T1}
	The vacuum (Casimir) energy for a spherical cavity became of high interest
	after the introduction of Hendrik
	Casimir’s proposal
	of an electron model in which the electrostatic repulsion is compensated for by an attraction from the vacuum 
	energy
	. In~1968,  Timothy Boyer~\cite{boye68-174-1764}
	showed that a conducting spherical shell leads to repulsion, a~result that was not regretted by everyone. Indeed, Casimir himself 
	later expressed satisfaction \cite{bord98}
	~that this outcome ruled out the hoped-for stabilization of the electron model by vacuum~fluctuations.
	
	Interest in this topic also came from the bag model in QCD (quantum chromodynamics).
	In~the MIT (Massachusetts Institute of Technology) 
	bag model, quarks and gluons are confined within a spherical cavity with appropriate boundary conditions. The~vacuum (Casimir) energy of the confined fields provides an essential contribution to the total hadron energy and plays a key role in the phenomenological stabilization of the bag radius. Early calculations of the zero-point energy of gluon fields in a spherical bag 
	by Kimbal 
	Milton~\cite{milt80-22-1441} and others,
	as~well as more detailed studies for massive Dirac fields in the MIT bag model~\cite{eliz98-31-1743}, highlighted the importance of proper renormalization and the dependence on the bag radius, thereby directly motivating detailed studies of massive and massless fields in spherical~geometries.
	
	The relation between vacuum (Casimir) energy and the cosmological constant ($\Lambda$) has also attracted significant attention. In~quantum field theory, the~vacuum energy density is expected to contribute to $\Lambda$ via the Einstein equations, yet naive estimates yield values many orders of magnitude larger than the observed cosmological constant (the famous ``cosmological constant problem''). When embedding quantum fields  in a gravitational background, renormalization of the vacuum energy in curved spacetime generally requires additional counterterms quadratic in the curvature scalars (such as $R^2$ and $R_{\mu\nu}R^{\mu\nu}$)
	(where the Greek letters take four space-time indexes), 
	which are not present in Einstein's general relativity~\cite{birr82b}. Even after these subtractions, 
	a~relatively
	small finite remnant vacuum energy typically remains, contributing to an effective cosmological constant. This issue has been discussed extensively in the literature, including in works by 
	Viatcheslav 
	Mukhanov and collaborators on dynamical approaches to the small cosmological constant~\cite{arme00-85-4438} and by 
	Andreas Wipf and collaborators in the context of zeta-function regularization and Casimir energy in curved 
	backgrounds~\cite{blau88-209-209}.
	
	Boyer’s calculation~\cite{boye68-174-1764}
	was complicated and benefited from several compensations of ultraviolet divergences between the interior and 
	exterior of the shell and between the modes of the electromagnetic field. This motivated studies of rectangular cavities (or 
	boxes), which provide another fundamental geometry for studying vacuum energies, often serving as a simpler setting for mode 
	summation and regularization techniques. Early calculations for massless and massive scalar fields (as well as electromagnetic 
	fields) in rectangular geometries were performed by Serge\u{\i} Mamaev and 
	Nikolay Trunov~\cite{mama79-22-966,mama79-38-228} in the late 1970s. Here, 
	it was sufficient to subtract the Minkowski-space contribution. These works explored the dependence of the Casimir energy on the cavity dimensions, revealing that the sign of the energy can exhibit a smooth transition from repulsive to attractive contributions depending on the aspect ratios of the~box.
	
	In the mid-1990s, there was a revival of interest in the spherical configuration in the context of heat kernel coefficients on manifolds with 
	boundaries
	. New methods were applied---notably, zeta-functional regularization, with its analytical beauty. The~topic became much broader, 
	including temperature effects, dependence on geometry and topology, and~effects on material properties in view of ever more precise 
	experiments (see, for~instance,~Refs. \cite{BKMM} or~\cite{milton01}).
	
	In 1997, in~Ref. \cite{bord97-56-4896}, the~Casimir energy of a massive field in a spherical cavity was calculated for the interior alone for the first time. This 
	calculation
	was of special interest, as~the mentioned compensations of divergences are absent here. However, the~mass provided a 
	novel
	, physically well-founded means of fixing the arbitrariness of the renormalization. One needs to demand that the vacuum energy vanish for 
	relatively large mass or, equivalently, for a small enough Planck constant. Regrettably, in~that paper, there is a sign error in one formula, causing the renormalized vacuum energy for 
	relatively large mass to vanish with the 
	incorrect
	sign. This resulted in a disagreement between Equation~(9.2) and Figure~9.2 in~Ref.~\cite{BKMM}, which was only recently 
	pointed out to us\footnote{Private communication by Jean Alexandre.}.
	This remark caused a reconsideration of the calculations, to~which the current paper is devoted. It turned out that some sums 
	left in Ref.~\cite{bord97-56-4896} for numerical evaluation can be expressed in terms of~polylogarithms.

	The present paper provides a detailed reconsideration of the calculations in~Ref. \cite{bord97-56-4896}. 
	We give
	the paper an introductory part (Section \ref{T2}), which makes it self-contained. We included the relation to the heat kernel 
	coefficients and a discussion of the renormalization. In~Section~\ref{T3}, we present the calculation of the related zeta function 
	in full detail. In~the last Section \ref{T4}, we present a discussion of the results.
	\section{Basic Formulas for Vacuum Energy and Zeta~Function}\label{T2}
	The (one-loop) vacuum energy of a quantum field can be represented as the half sum of the frequencies 
	$\om_J$ 
	of the field modes, i.e.,
	\eq{2.1}{ E_0(s) &= \frac{\hbar }{2}\sum_J\om_J^{1-2s}%,
	}
	which follows from a Klein--Gordon equation, i.e.,
	\eq{2.2}{ \left( - \Delta+\frac{c^2 m^2}{\hbar^2}  \right)\phi_J(x) = \frac{\om_J^2}{c^2}\,\phi_J(x),
	}
	with $m$ the mass of the particle, $\hbar$ the reduced Planck constant, and $c$ the speed of light  and $\phi_J$ the scalar field,   together with a boundary condition.  
	In the following we move on to the system of units $\hbar=c=1$.
	We use the zeta-functional 
	regularization with the $s$ parameter, which must be taken as zero after renormalization. For~simplicity, we do not introduce the $\mu$ parameter, which is commonly used to describe the arbitrariness that comes with regularization.
	For the spherical geometry, separating the variables, i.e.,
	$\phi_J(x)=e^{-i\om t+il\varphi}\phi_l(r)$, the~equation turns into
	\eq{2.3}{  \left(
		-\frac{\pa^2}{\pa r^2}-\frac{1}{r}\frac{\pa}{\pa r}+\frac{l(l+1)}{r^2}
		+m^2  \right) \phi(r) &= \om^2\,\phi_l(r),
	}
	and we take the boundary condition (Dirichlet) as
	\eq{2.4}{ \phi_l(R)=0.
	}
	Equation \Eq{2.3} 
	is a Bessel equation, and it has solutions of
	\eq{2.5}{ \phi_J(r) &= J_{l+1/2}(\om_{l,n} r), %&
		\quad \mbox{with}\quad %&& 
		\om_{l,n}%&
		=
		\sqrt{\left(\frac{j_{l,n} }  {R}\right)^2+m^2},
	}
	where $j_{l,n}$ represents the zeros of the Bessel function $J_n$.
	The vacuum energy \Eq{2.1} takes the more special
	form of
	\eq{2.6}{  E_0(s) &= \frac{1}{2}\sum_{n=1}^\infty\sum_{l=0}^\infty
		(2l+1)\om_{l,n}^{1-2s}.
	}
	It is convenient to introduce the zeta function, i.e.,
	\eq{2.7}{  \zeta(s) &= \sum_{n=1}^\infty\sum_{l=0}^\infty
		(2l+1)\om_{l,n}^{-2s},
	}
	and the vacuum energy can be expressed as
	\eq{2.8}{  E_0(s) &= \frac{1 }{2}\zeta(s-1/2).
	}
	As a matter of fact
	, 
	$\zeta(s)$ is the zeta function of the Laplace operator of the three-dimensional ball with Dirichlet boundary conditions~\cite{seel67b}. It is a meromorphic function of $s$ with simple poles on the real axis for $s\le3/2$. This zeta function can be represented as an integral, i.e.,
	\eq{2.9}{  \zeta(s) &= \int_0^\infty\frac{dt}{t}\frac{t^s}{\Ga(s)}\,K(t)\,e^{-t m^2},
	}
	where
	\eq{2.10}{  K(t) &= \sum_{n=1}^\infty\sum_{l=0}^\infty e^{-t(j_{l,n}/R)^2}
	}
	is the related heat kernel and   $\Gamma(s)$ is the gamma function. 
	For $t\to0$, $K(t)$ is known to have an asymptotic expansion (heat kernel expansion), i.e.,
	\eq{2.11}{ K(t) &= \frac{1}{(4\pi t)^{3/2}}\sum_{n\ge 0}a_nt^n,
	}
	where $a_n$ represents the {heat kernel coefficients}. 
	The expansion \Eq{2.11} represents    
	a semiclassical expansion for 
	considerably large values of $m$ and 
	small enough 
	values of $\hbar$ (if we recover the fundumental constants) with integer and, due to the boundary, half-integer indices. The~heat kernel coefficients \Eq{2.11} 
	provide a universal language for describing the ultraviolet divergences of the vacuum energy, since these appear in the integral representation at $t\to0$. 
	In~the case of three-dimensional space under consideration,  
	$n=0,\frac12,\dots,2$ contribute to the ultraviolet~divergences.
	
	Inserting the heat kernel expansion into the zeta function \Eq{2.9} and the vacuum \mbox{energy \Eq{2.6},} we define
	the divergent part of the zeta function as
	\eq{2.12}{ \zeta^{\rm div}(s) &=	\sum_{n\le 2}a_n 
		\int_0^\infty\frac{dt}{t}\frac{t^{s+n-3/2}}{(4\pi)^{3/2}\Ga(s)}\,e^{-tm^2}%,
		\\\nn & = 	\sum_{n\le 2}a_n
		\frac{\Ga(s+n-3/2)}{(4\pi)^{3/2}\Ga(s)}\,m^{3-2n-2s},
	}
	and the divergent part of the vacuum energy as
	\eq{2.13}{ E^{\rm div}(s) &=\frac{1 }{2}\zeta^{\rm div}(s-1/2).
	}
	Explicitly written, 
	the expansion \Eq{2.13}
	reads 
	\eq{2.14}{ E^{\rm div}(s) &=
		-\frac{m^4\, a_0}{64\pi^2}
		\left[\frac{1}{s}-\frac12\left(1+\ln%\left(
		\frac{m}{2}%\right)
		\right)
		\right]
		-\frac{m^3\, a_{1/2}}{24\pi^{3/2}}
		+\frac{m^2\, a_1}{32\pi^2}\left[
		\frac{1}{s}-\left(1+2\ln%\left(
		\frac{m}{2}%\right)
		\right)
		\right]
		\\ \nn	&~~~~
		+\frac{m\, a_{3/2}}{16\pi^{3/2}}
		-\frac{  a_{2}}{32\pi^2}
		\left[
		\frac{1}{s}-2\left(1+ \ln%\left(
		\frac{m}{2}%\right)
		\right)
		\right],
	}
	with 
	the coefficients 
	\eq{2.15}{ a_0&=\frac{4}{3}\pi R^3,& a_{1/2}&=-2\pi^{3/2}R^2, &a_1&=\frac{8}{3}\pi R,
		& 
		a_{3/2}&=-\frac{1}{6}\pi^{3/2},
		& 
		a_2&=-\frac{16}{315}\frac{\pi}{R}.
	}
	The 
	formulas \Eq{2.14}--\Eq{2.15}  
	reproduce Equation~(13) of Ref.~\cite{bord97-56-4896} up to 
	$a_2$, which regrettably has the opposite sign there. In~Equation~.
	\Eq{2.14}, the arbitrary $\mu$ factor, which makes the 
	arguments of the logarithms dimensionless, is omitted (set to $\mu=1$). 
	The heat kernel coefficients \Eq{2.11} can  be found elsewhere
	---notably, in Ref.~\cite{vass03-388-279}.
	
	Now, $ E^{\rm div}(s)$ \Eq{2.14} 
	carries the pole contribution in $s$, and the limit of $s\to0$, i.e.,
	\eq{2.16}{ E^{\rm ren} &= \lim\limits_{s\to0}\left(E_0(s)-E^{\rm div}(s)\right),
	}
	is finite and defines the renormalized vacuum 
	energy
	. The radius of the sphere 
	$R$
	~enters into the expressions for energy as 
	$E \propto
	R^{-1} f(m R)$, where $f$ is dimensionless. Furthermore, in what follows, we assume $R=1$ unless otherwise stated. The~dependence on $R$ can be restored in the formulas by substituting $E\to E/R$ and $m\to m R$.
	
	There is an arbitrariness in the renormalization procedure. In
	Equations~\Eq{2.16},
	not only is the (divergent) pole part subtracted but~also some finite part including the terms depending on  
	$\mu$, if~restored. 
	As~follows from the heat kernel expansion \Eq{2.11},
	it has the property such that $E^{\rm ren}$ \Eq{2.16} 
	~vanishes in the large
	mass limit, i.e.,
	\eq{2.17}{  \lim\limits_{m\to\infty} E^{\rm ren}=0.
	}
	Equation \Eq{2.17} represents 
	a physical requirement, since an infinitely heavy field should not have quantum fluctuations. Upon~restoring the dimensions, only non-negative powers of $\hbar$ are left, since the mass enters in the combination of $m/\hbar$. 
	Let us
	mention that this condition is not applicable to a massless field---the~electromagnetic one, for instance 
	(however, 
	see a remark we make in concluding Section \ref{T5}).
	
	As for the relation between the zeta function and the heat kernel coefficients, let us
	mention two 
	quite straightforward 
	checks, which, in the 1990s, were used to calculate a number of coefficients resulting from the~boundary. 
	
	One of the checks 
	is related to Equation~\Eq{2.12} (second line). Truncating the sum after $n=2$, the~poles in $s=-1/2$ give a direct relation between the powers of the mass 
	(note:
	~this is for $m\to\infty$) and the coefficients with integer index, whereas the coefficients with half-integer index follow from the regular 
	parts. A~more explicit expression is Equation~ \Eq{2.14} (with Equation 
	\Eq{2.13}) when the $\zeta^{\rm div}$ pole and regular parts are known in $s=-1/2$ in the 
	large
	-mass~limit
	.
	
	For the other check, one needs to know the zeta function as a function of $s$ but~at $m=0$, which can be calculated 
	much more straightforwardly (see~\cite{bord96-37-895}). 
	Consider the pole part of $\Ga(s)\zeta(s)$ with \Eq{2.12} at $m=0$. Integrating over $t$ with substitution of $K(t)$ at the point $t\to0$ by its asymptotic expansion \Eq{2.11} yields the formula
	\eq{2.19}{  a_n &= \operatorname*{Res}\limits_{s=3/2-n} (4\pi)^{3/2}
		\Ga(s)\zeta(s)  |_{m=0},
	}
	which allows for a straightforward calculation of the coefficients if the zeta function is~known.
	
	With \Eq{2.16}, one has an expression for the renormalized vacuum energy. A way to perform the limit of $s\to0$ remains to be described. For~this one, 
	let us 
	define a suitable {\it asymptotic
		part}
	, i.e., $E^{\rm as}(s)$, which is subtracted and added back to split
	\eq{2.20}{ E^{\rm ren} &= E^{\rm num}+E^{\rm an}
	}
	with
	\eq{2.21}{ E^{\rm num} %&
		=
		\lim\limits_{s\to 0} %\left
		\Big(E_0(s)-E^{\rm as}(s)%\right
		\Big)%,
		\quad
		%& 
		\text{and} %& 
		\quad 
		E^{\rm an}%&
		=
		\lim\limits_{s\to 0} \left( E^{\rm as}(s)-E^{\rm div}(s)\right).
	}
	To define $E^{\rm as}(s)$, we use the representation of the zeta function \Eq{3.5} (see just
	below) and insert the uniform asymptotic expansion of the modified Bessel function for both large enough argument and index. With~the 
	newly defined  variable, i.e.,
	\eq{2.22}{ z&=\nu \xi,
	}
	the modified Bessel function 
	reads
	\eq{2.23}{ I_\nu(\nu z) &\approx %\sim 
		\phi^{\rm as}\equiv\frac{e^{\nu\eta(z)}}
		{\sqrt{2\pi \nu}(1+z^2)^{1/4}}
		\sum_{k=-1}^3\frac{u_k(t)}{\nu^k},
		\\\nn	 \text{with}\quad t&=\frac{1}{\sqrt{1+z^2}}%,
		\quad	\text{and}\quad  \eta(z)=\sqrt{1+z^2}+\ln\frac{z}{1+{\sqrt{1+z^2}}},
	}
	along with the known Debye polynomials ($u_k(t)$; see~\cite{NIST:DLMF})
	.  This gives rise to the definition of
	\eq{2.24}{ \zeta^{\rm as}(s) &= 2\frac{\sin(\pi s)}{\pi}
		\sum_{l=0}^\infty \nu \int_{m/\nu}^\infty dz
		\left((\nu z)^2-m^2\right)^{-s} \frac{\pa}{\pa z}
		\ln\phi^{\rm as}.
	}
	In analogy to Equation \Eq{2.13}, Equation \Eq{2.24} defines $E^{\rm as}(s)$ as used in Equation \Eq{2.21}.
	
	The merit of the  definition \Eq{2.24}  
	is twofold. For~$E^{\rm num}$, one gets the representation expressed as
	\eq{2.25}{  E^{\rm num} &=
		-\frac{1}{\pi}\sum_{l=0}^\infty \nu \int_{m/\nu}^\infty dz
		\sqrt{(\nu z)^2-m^2} \frac{\pa}{\pa z}
		\left(	\ln \phi-\ln\phi^{\rm as}\right).
	}
	Here, one may set
	$s=0$ under the signs of summation and integration. Due to the properties of the uniform asymptotic expansion \Eq{2.23}, both 
	the integration and the subsequent summation do~converge.
	
	For the $E^{\rm an}$ term in Equation \Eq{2.20}, it is worth to note, with Equation \Eq{2.12}, that
	\eq{2.26}{ \zeta^{\rm an}(s) &=\zeta^{\rm as}(s)-\zeta^{\rm div}(s).
	}
	The asymptotic zeta function ($\zeta^{\rm as}(s)$) can be calculated, to a large extent, analytically---at~least the pole part in $s$---which allows for direct analytic continuation in $s$. After~that, the pole parts in 
	Equation \Eq{2.26} cancel, and one is left with a finite~expression.
	\section{Calculating the Zeta~Function}\label{T3}
	The zeta function was defined in Equation \Eq{2.7}. With~Equation\Eq{2.5}, the zeta fuction takes the form of
	\eq{3.1}{  \zeta(s) &= 2\sum_{l=0}^\infty \,\nu\sum_n \left(j_{l,n}^2+m^2\right)^{\frac12-s}.
	}
	Here, we use the notation 
	\eq{3.2}{ \nu &=l+\frac12,%.
	}
	which is convenient for what follows. The~sum over $n$ in Equation \Eq{3.1} 
	is taken 
	over all Bessel zeros. According to Equation
	~\Eq{3.1}, the~zeta function is defined for $s>3/2$ and the task is its analytic continuation to lower $s$ values---$s=-1/2$, for~instance.
	There are a few strategies to proceed. One way 
	is to perform the orbital sum first and to make the continuation in the remaining sum. Another 
	is to use the known uniform asymptotic expansion of the zeros. But~the most convenient way is to transform the sum over $n$ into an integral 
	and to deform the integration path towards imaginary frequencies. We apply this~method.
	One needs  
	a {\it  mode-generating function}, 
	i.e.,~a meromorphic function whose zeros are the Bessel zeros. 
	A~comparably
	~simple
	choice is 
	\eq{3.3}{  \Phi(\la) &=\la^{-\nu}J_\nu(\la),
	}
	such that $\Phi(\la)=0$ has as its solution, i.e., $\om=j_{l,n}$. The~$\la^{-\nu}$ factor makes $\Phi(\la)$ regular at the~origin.
	Using the Cauchy theorem and~the mode-generating function, the~sum over $n$ in Equation \Eq{3.1} can be rewritten 
	as
	\eq{3.4}{  \zeta(s) &=\sum_{l=0}^\infty \nu \int_\ga\frac{d\la}{2\pi i}
		\left(%[
		\la^2+m^2\right)%]
		^{-s}
		\frac{\pa}{\pa \la} \ln \Phi(\la),
	}
	where the integration path ($\ga$) encircles the relevant zeros of $\Phi(\la)$. In~the next step, one turns the integration path towards the imaginary axis, and using some analytic properties, one arrives at
	\eq{3.5}{  \zeta(s) &=
		2\frac{\sin(\pi s)}{\pi}\sum_{l=0}^\infty \nu \int_m^\infty d\xi
		\left(\xi^2-m^2\right)^{-s}\frac{\pa}{\pa \xi}\ln\Phi(i\xi),
	}
	where, in the case considered,
	\eq{3.6}{ \Phi(i\xi) &= \xi^{-\nu}I_\nu(\xi)
	}
	(up to an irrelevant constant).
	
	The main part of this study
	is the calculation of the asymptotic part. With~Equations \Eq{2.24} and \Eq{2.23}, we define
	\eq{3.7}{ \zeta^{\rm as}(s) &=\sum_{k=-1}^3 A_k(s)
	}
	with
	\eq{3.8}{   A_{k}(s)&=2\frac{\sin(\pi s)}{\pi}\sum_{l=0}^\infty   \nu^{1-k}
		\int_{m/\nu}^\infty dz\  \Big((\nu z)^2-m^2\Big)^{-s}\pa_z D_k(t),	%\qquad t=\frac{1}{\sqrt{1+z^2}}.
	}
	where $\pa_z \equiv \pa/\pa_z$.
	The  polynomials 
	$D_k(t)$
	~follow from the re-expansion of Equation \Eq{2.23} for large enough $\nu$ values such that  we define
	\eq{3.9}{ \ln\phi^{\rm as} &= \sum_{k=-1}^3\frac{D_k(t)}{\nu^k}.
	}
	Note that just the expression \Eq{3.9} must be used for $ \ln\phi^{\rm as} $ 
	in Equation \Eq{2.25}. Explicit expressions for 
	$D_k(t)$,
	~$k=1,2$, and $3$, are 
	displayed in Sections \ref{T33}--\ref{T35} below; {the first two expressions read
		\eq{3.9a}{D_{-1}(t) %&
			= \frac{1}{t}-\arctan%(
			t%)%,
			\quad\text{and}\quad %&
			D_0(t)%&
			=\frac12\ln%[
			\frac{t}{2\pi\nu}%]
			.
	}} 
	It is worth mentioning that for $k=1,2,$ and $3$, 
	$D_k$ are polynomials in $t$.
	
	In the following Sections \ref{T31}--\ref{T35}, we recalculate the individual $A_k(s)$.
	\subsection{$A_{-1}$}
	\label{T31}
	For $A_{-1}$, we use the representation obtained in Ref.~\cite{bord97-56-4896} (see Equation~(A6) in 
	Ref.~\cite{bord97-56-4896}):%.
	\eq{3.1.1}{ A_{-1}(s)&= \frac{\Ga(s-1/2)}{2\sqrt{\pi}\,\Ga(s)}
		\sum_{l=0}^\infty \nu
		\int_0^1	dx \,x^{s-1}
		(\nu^2+x m^2)^{1/2-s} ,%\q
		\quad %(
		\nu=l+1/2.%)
	}
	In the following, we use the more convenient notation 
	\eq{3.1.1a}{s=\ep-\frac12
	}
	and need the continuation to $\ep=0$. First, let us 
	rewrite 
	Equation \Eq{3.1.1} as
	\eq{3.1.2}{  A_{-1}(\ep)&= \frac{\Ga(\ep-1)}{2\sqrt{\pi}\,\Ga(\ep-1/2)}
		\sum_{l=0}^\infty \int_0^1
		dx \,x^{\ep-3/2} \nu^{3- 2\ep}\left(1+\frac{x\,m^2}{\nu^2}\right)^{1-\ep}.
	} 
	One obtains 
	the asymptotic part of Equation \Eq{3.1.2},
	expanding the expression in parentheses, raised to the power ${1-\ep}$, into~a series
	for $\nu\to\infty$, i.e,
	\eq{3.1.3}{\left(1+\frac{x\,m^2}{\nu^2}\right)^{1-\ep} \to  a_s &\equiv 1+\frac{(1-{\ep}) m^2 x}{\nu ^2}+
		\frac{({\ep}-1) {\ep} m^4 x^2}{2 \nu ^4}.
	} 
	Then, we add and subtract $a_s$ from the integrand and split Equation \Eq{3.1.2} as follows:
	\eq{3.1.4}{ A_{-1} &= A_{-1}^{\rm as}+A_{-1}^{\rm sub},
	}
	where $A_{-1}^{\rm as}$ and $A_{-1}^{\rm sub}$
	result%s 
	~from the subtraction of $a_s$
	, i.e.,
	\eq{3.1.5}{ A_{-1}^{\rm as} &= \frac{\Ga(\ep-1)}{2\sqrt{\pi}\,\Ga(\ep-1/2)}
		\sum_{l=0}^\infty \int_0^1
		dx \,x^{\ep-3/2} \nu^{3-2\ep} a_s,
		\\\nn
		A_{-1}^{\rm sub} &= \frac{\Ga(\ep-1)}{2\sqrt{\pi}\,\Ga(\ep-1/2)}
		\sum_{l=0}^\infty \int_0^1
		dx \,x^{\ep-3/2} \nu^{3-2\ep}
		\left[ \left(1+\frac{x\,m^2}{\nu^2}\right)^{1-\ep}-a_s\right]~~~.
	} 
	In $A_{-1}^{\rm as}$, the~summation is explicit in terms of zeta functions and so is also the subsequent $x$-integration. Finally, the~expansion in $\ep$ delivers a pole and regular terms, i.e., 
	\eq{3.1.6}{A_{-1}^{\rm as}&=A_{-1}^{\rm as(p)}/\ep+A_{-1}^{\rm as(reg)}+{\mathcal O}(\ep),  
		\\\nn
		A_{-1}^{\rm as(p)} &= \frac{-80 m^4+40 m^2+7}{1920 \pi },
		\\\nn 	A_{-1}^{\rm as(reg)} &=
		\frac{-3m^2 (-1+12\zeta'(-1)+3)-2m^4 (-4+3 \gamma +9 \log 2)}{72 \pi} \\\nn
		& +	\frac{1680 \zeta '(-3)+7+12 \log 2}{1920 \pi }\,%~~~
		,%.
	} 
	where the prime denotes the $s$-derivative.
	In $A_{-1}^{\rm sub}$, we expand in $\ep$ since the summation is convergent. The~pole in the prefactor cancels, and  one arrives at
	\eq{3.1.7}{  A_{-1}^{\rm sub}&= 	\sum_{l=0}^\infty \int_0^1 dx	\frac{	\nu ^3}{4 \pi  x^{3/2}} \,
		\left[\frac{m^4 x^2}{2 \nu ^4}+\frac{m^2 x}{\nu ^2}-\left(\frac{m^2 x}{\nu ^2}+1\right) \log \left(\frac{m^2 x}{\nu
			^2}+1\right)\right]+{\mathcal O}(\ep)~~~.
	} 
	The integration over $x$ is also straightforward
	:
	\eq{3.1.8}{   A_{-1}^{\rm sub}&= \frac{1}{12\pi}	\sum_{l=0}^\infty
		\left[
		\frac{m^4}{\nu}+18m^2\nu-24m\nu^2\arctan%\left(
		\frac{m}{\nu}%\right)
		+6 \nu (\nu^2-m^2)\ln\left(1+\frac{m^2}{\nu^2}\right)
		\right] . 
	}
	The sums are more demanding but~can be done.  One arrives at
	\eq{3.1.9}{  A_{-1}^{\rm sub} &=
		\frac{m^4}{12 \pi }
		(\gamma +\log %(
		4%)
		)-12 m^2  \psi ^{(-2)}\left(\frac{1}{2}\right)
		-12 m i %( 
		\psi ^{(-3)}\left(\frac{1}{2}-i m\right)	\\
		\nn&+12 m i\psi ^{(-3)}\left(i m+\frac{1}{2}\right)
		-36 \psi ^{(-4)}\left(\frac{1}{2}-i m\right)-36 \psi
		^{(-4)}\left(i m+\frac{1}{2}\right)+72 \psi ^{(-4)}\left(\frac12\right)~~~.
	} 
	Here, $\psi^{(n)}(z)$ is a polygamma function.
	Let us mention that expression \Eq{3.1.9} is~real.
	Combining
	parts $A_{-1}^{\rm as}$ and $A_{-1}^{\rm sub}$, for $A_{-1}$ \Eq{3.1.1}, one gets a~pole part that results only from $A_{-1}^{\rm as(p)}$ \Eq{3.1.6} and~a regular part that comprises $A_{-1}^{\rm as(reg)}$ \Eq{3.1.6} and~$A_{-1}^{\rm sub}$ \Eq{3.1.9}, 
	i.e.,
	
	\eq{3.1.10}{ A_{-1}^{p} &=  \frac{-80 m^4+40 m^2+7}{1920 \pi }\,%~~~
		,
		\\\nn
		A_{-1}^{\rm reg} &= \frac{1}{5760 \pi }
		\bigg[
		-2880 m^2 \zeta'(-1)+5040 \zeta '(-3)+640 m^4-480 m^4 \log 2-480 m^2
		\\\nn &	~~~
		-5760 m^2 \psi ^{(-2)}\left(\frac{1}{2}\right)-5760 i m
		\psi ^{(-3)}\left(\frac{1}{2}-i m\right)
		+5760 i m \psi ^{(-3)}\left(i m+\frac{1}{2}\right)
		\\ \nn &~~~
		-17280 \psi
		^{(-4)}\left(\frac{1}{2}-i m\right)	
		-17280 \psi ^{(-4)}\left(i m+\frac{1}{2}\right)	
		\\\nn &~~~+21+36 \log 2+34560 \psi
		^{(-4)}\left(\frac{1}{2}\right)    \bigg]~~~.
	}  
	The expansion of the expression \Eq{3.1.10}
	for $m\to\infty$ reads
	\eq{3.1.11}{ 	A_{-1} ^{\rm reg} &\approx%\sim
		\frac{m^4 (4 \log {\frac{m}{2}} %(m/2)
			+1)}{48 \pi }-\frac{m^2 (2 \log {\frac{m}{2}}%(m/2)
			+1)}{48 \pi }
		-\frac{7 (\log {\frac{m}{2}}%(m/2)
			+1)}{960
			\pi }+{\mathcal O}\left(\frac{1}{m^2}\right)~.
	}  
	We also consider the expression for $A_{-1}$ at $m=0$. From Equation 
	\Eq{2.23} by direct integration over $x$ and summation over $l$ one obtains 
	\eq{3.1.12}{ A_{-1}^{m=0} &= \frac{\left(2^{2 s-2}-1\right) \zeta \Big(2 (s-1)\Big) \Gamma \left(s-\frac{1}{2}\right)}{2 \sqrt{\pi } \Gamma 
			(s+1)}~.
	} 
	\subsection{$A_{0}$}
	\label{T32}
	Here, we follow Equation (A10) 
	from Ref.~\cite{bord97-56-4896}, i.e.,
	\eq{3.2.1}{  A_0 &= \frac{m^3}{6}-\int_0^m \frac{d\nu\,\nu}{1+\exp(2\pi \nu)}
		\sqrt{m^2-\nu^2}~~~,
	}
	which was obtained after integration over $z$  in Equation 
	\Eq{3.8} with $k=0$ by applying the Abel--Plana formula to the remaining summation. This contribution does not have a pole term and 
	can be used as is. To~conform the notations in this Section, we define
	\eq{3.2.2}{ A_0^{p} &=0,
		\\\nn A_0^{\rm reg}&= A_0~~~.
	}
	The expansion for large $m$ values also follows immediately, i.e.,
	\eq{3.2.3}{  A_0^{m\to\infty} &\approx%\sim
		\frac{m^3}{6}-\frac{m}{48}+\mathcal{O}%O
		\left(\left(\frac{1}{m}\right)^1\right)~~~.
	}
	For $m=0$, one  returns  
	to the initial expression \Eq{3.8} with $D_0=-\frac14\ln(1+z^2)$, which can be integrated and summed up directly, resulting in
	\eq{3.2.4}{ A_0^{m=0} &= -\frac{1}{2} \left(2^{2 s-1}-1\right) \zeta (2 s-1)~~~.
	}
	\subsection{$A_{1}$}
	\label{T33}
	For $A_1$ (and the following $A_k$, $k=2,3$),
	we start from Equation
	\Eq{3.8}, i.e.,
	\eq{3.3.1}{ A_1 &= 2\frac{\sin(\pi s)}{\pi}\sum_{l=0}^\infty
		\int_{m/\nu}^\infty dz \ \Big((\nu z)^2-m^2\Big)^{-s}\pa_z D_1(t)\,%~~~
		,
		%\qquad t=\frac{1}{\sqrt{1+z^2}}~~~,
	}
	with  $D_1(t)$ defined according to Equation  \Eq{3.9}, i.e.,
	\eq{3.3.2}{ D_1(t)&= \frac{(3-5t^2)t}{24}~~~.
	}
	It is worth
	to first integrate over $z$ in Equation \Eq{3.3.1}, setting  $s=\ep-1/2$. As~a result, one gets  
	\eq{3.3.3}{  A_1 &= c_1\sum_{l=0}^\infty   (w_1+w_2), \quad  c_1= -
		\frac{\cos (\pi \ep) \Gamma \left(\frac{3}{2}-{\ep}\right)
			\Gamma ({\ep})}{12 \pi ^{3/2}}~~~,
	}
	where
	\eq{3.3.6}{
		w_1=-3 \nu ^{1-2\ep} \left(\frac{m^2}{\nu ^2}+1\right)^{-\ep}%, &&
		\quad\text{and}\quad w_2=10\ep \nu ^{1-2\ep} \left(\frac{m^2}{\nu ^2}
		+1\right)^{-\ep-1}\,%~~~
		.
	}
	The sum over $l$ in Equation 
	\Eq{3.3.3} can be calculated in the following way. We define the asymptotic parts of $w_1$ and $w_2$ as
	\eq{3.3.7}{ w_1^{\rm as}&=-3 \nu ^{1-2\ep}
		\left(1-\frac{\ep m^2}{\nu ^2}+\mathcal{O}%O
		\left( \nu^{-3 }\right)\right)\,%~~~
		,
		\\\nn
		w_2^{\rm as}&=10 \ep \,
		\nu ^{1-2\ep}
		\left(1-\frac{  ({\ep}+1) m^2}{\nu ^2}+\mathcal{O}%O
		\left(\nu^{-3}\right)\right)\,%~~~
		,
	}
	and then the corresponding regular 
	parts in Equation \Eq{3.3.3},
	already multiplied by $c_1$ \Eq{3.3.3}, as
	\eq{3.3.8}{ w_1^{\rm sub} = c_1 (w_1-w_1^{\rm as}) %,& \q
		\quad\text{and}\quad 
		w_2^{\rm sub} = c_1( w_2-w_2^{\rm as})\,%~~~
		.  
	}
	The sums in the regular parts converge, and setting $\ep=0$, one obtains
	\eq{3.3.9}{  w_1^{\rm sub} &= \frac{m^2-\nu ^2 \log
			\left(\frac{m^2}{\nu ^2}+1\right)}{8 \pi  \nu }\,%~~~
		,
		\\\nn
		w_2^{\rm sub} &= -\frac{5}{12 \pi \nu} \frac{m^4}{   m^2  +  \nu ^2}\,%~~~
		. 
	}
	According to the split of $w_i$ into regular and asymptotic parts  in Equation \Eq{3.3.3},
	we split $A_1$ \Eq{3.3.1} into~%
	\eq{3.3.10}{ A_1=A_1^{\rm sub}+A_1^{\rm as},
	}
	where
	\eq{3.3.11}{  A_1^{\rm sub} &= \sum_{l=0}^\infty (w_1^{\rm sub}+w_2^{\rm sub})%~~~,& 
		\quad\text{and}\quad
		A_1^{\rm as}%&
		=c_1\sum_{l=0}^\infty ( w_1^{\rm as}+w_2^{\rm as})\,%~~~
		.
	}
	In the regular terms, the~summation is explicit, i.e.,
		\eq{3.3.12}{ A_1^{\rm sub}&=\frac{1}{24\pi}\bigg%\Bigl
			[9\zeta'(-1) -\frac{3}{4}+m(3 i -5m) \psi^{(0)}\left(\frac{1}{2}-i m\right)- m(3 i +5m) 
			\psi^{(0)}\left(i m+\frac{1}{2}\right) \bigg.%\Bigr.
			\\\nn \bigg.%\Bigl.
			&-7 m^2( \gamma +2 \log 2)
			-\frac{5}{4} \log 2
			+3 \psi ^{(-2)}\left(\frac{1}{2}-i
			m\right)+3 \psi ^{(-2)}\left(i m+\frac{1}{2}\right)-\frac{3 \log %(
				\pi %)
			}{2}\bigg%\Bigr
			]\,%~~~
			.
		}
	In the asymptotic part 
	$A_1^{\rm as}$
	, the~summations are in terms of zeta functions. There are pole terms in addition to regular terms, i.e., $A_1^{\rm as}=A_1^{\rm as(p)}+A_1^{\rm as(reg)}$:
	\eq{3.3.13}{ A_1^{\rm as(p)}&=
		\frac{28 m^2+1}{192 \pi  {\ep}}\equiv \frac{A^{\rm p}_1}{\ep}\,%~~~
		,
		\\\nn 
		A_1^{\rm as(reg)}&=
		\frac{-9\zeta'(-1) -2}{72 \pi }+\frac{m^2 (-2+7 \gamma +21 \log 2
			)}{24 \pi }+{\mathcal O}\left(\ep^1\right)\,%~~~
		.
	}
	Combining 
	the contributions of Equations \Eq{3.3.12} and \Eq{3.3.13}, one arrives 
	at
	\eq{3.3.14}{ 
		A_1 &=\frac{1}{\epsilon}A_1^{p}+A_1^{\rm reg}\,%~~~
		,\\ 
		A_1^{p} &=\frac{7 m^2}{48 \pi }+\frac{1}{192 \pi }\,%~~~
		,
		\\\nn 
		A_1^{\rm reg}&=A_1^{\rm sub}+A_1^{\rm as(reg)}=-\left.\frac{1}{48 \pi}\right\{\frac{17}{6} + 2 m^2 (2 - 7 \log 2) + \frac{5}{2} \log 2 -12 
		\zeta'(-1)  \\\nn &	~~
		+ 
		3 \log \pi 
		- 6 \left[%(
		\psi^{(-2)}\Bigl(\frac12 - i m\Bigr) +
		\psi^{(-2)}\Bigl( \frac12 + i m\Bigr)\right]%)
		+ 
		\\\nn &~~ \left.
		2 m \left[%(
		(5 m-3 i) \psi^{(0)}\Bigl(\frac12 - i m\Bigr) +(5 m+3 i)
		\psi^{(0)}\Bigl(\frac12 + i m\Bigr) \right]%)
		\right\}\,%~~~
		.	%
	}
	The expansion of $A^{\rm reg}_1$  at  $m\to\infty$ is also a strightforward  task and gives
	\eq{3.3.15}{ A_1^{\mathrm inf} &=
		-	\frac{7 m^2}{48 \pi} \left(2 \log \frac{m}{2}%(m/2)
		+1\right)
		-\frac{ \log \frac{m}{2}%(m/2)
			+1}{96 \pi
		}+\mathcal{O}%O
		\left(\frac{1}{m}\right)\,%~~~
		.
	}
	Not carrying out the summation in $A$
	~\Eq{3.3.12}, 
	together with $B$ 
	~\Eq{3.3.12},  we confirm $A_1$ in Equation~(A11) from~Ref.\cite{bord97-56-4896} up to the sign of the 
	$9\zeta'(-1)$-
	~term in the second line there, which must be~of the minus sign
	.
	The expression for $m=0$ can, as~above
	, be obtained from the initial expression, i.e., from Equation 
	\Eq{3.3.1} in this case, by~direct integration and summation and results in
	\eq{3.3.16}{ A_1^{m=0} &=
		-\frac{\left(2^{2 s}-1\right)}{6\pi^{3/2}} \left(5 s +1\right) \zeta \left(2 s\right) \cos \left(\pi  \Bigl(s+\frac{1}{2}\Bigr)\right) \Gamma (1-s) \Gamma
		\left(s+\frac{1}{2}\right)~~~. 
	}
	\subsection{$A_{2}$}
	\label{T34}
	Let us start from Equation \Eq{3.8} with $k=2$ and note that
	\eq{3.4.1}{ D_2(t) &= \frac{(1-6t^2+5t^4)t^2}{16}.%,& t&=\frac{1}{\sqrt{1+z^2}}
		%,& \nu&=l+\frac12.
	}
	Integrating over $z$ in Equation \Eq{3.8} results in
	\eq{3.4.3}{ A_2 &=-\frac{1}{128} \sum_{l=0}^\infty w\,%~~~
		,
	}
	where
	\eq{3.4.2}{ w&= 	(2\ep-1) \nu
		\left(m^2
		+\nu^2\right)^{-\ep
			-\frac{5}{2}} \left(8 m^4-8 (6\ep+1) m^2 \nu ^2+(2\ep-1) (10\ep+1) \nu ^4\right)\,%~~~
		. 
	}
	Let us 
	rewrite $w$ \Eq{3.4.2} in the form 
	\eq{3.4.4}{ w&=(\nu^2+m^2)^{-\ep} \nu \sum_{k=1}^3 p_k (m^2)^{3-k}(\nu^2+m^2)^{k-1-\frac{5}{2}}
	}
	{with} 
	$$p_1=-5\left(1-(2 \ep)^2\right) (3 + 2 \ep)~~~, \quad p_2=2 \left(1 - (2 \ep)^2\right)  (3 + 10 \ep)~~~, \quad p_3=(1-2\ep)^2(1+10 \ep)\,%~~~
	.$$
	Furthermore, using the notation 
	\eq{3.4.6}{ z=\sqrt{1+\frac{m^2}{\nu^2}},
	}
	one arrives at
	\eq{3.4.5}{ w&=(\nu^2+m^2)^{-\ep}\sum_{k=1}^3 \, p_k(z^2-1)^{3-k}z^{2k}\,%~~~
		.
	}
	Reordering the factors that are polynomial in $z$, i.e.,
	\eq{3.4.7}{  \sum_{k=1}^3 p_k(z^2-1)^{3-k}z^{2 k-7}
		= \frac{1}{z^5}\sum_{k=1}^3 q_kz^{2(k-1)}   
	}
	one arrives 
	at
	\eq{3.4.8}{ w=(\nu^2+m^2)^{-\ep}  \sum_{k=1}^3 \,q_k \,z^{2 k-7}%~~~
		,  
	}
	where  $q_1=p_1$, $q_2=-(2p_1+p_2)$, and $q_3=p_1+p_2+p_3$.
	Substituting Equation \Eq{3.4.8} into $A_2$ \Eq{3.4.3},
	we split  %
	\eq{3.4.9}{ A_2 &=A_2^{\rm as}+A_2^{\rm reg}, 
	}
	where
	\eq{3.4.10}{ A_2^{\rm as} %&
		=-\frac{1}{128} \sum_{l=0}^\infty (\nu^2+m^2)^{-\ep}  \sum_{k=1}^3 \,q_k%,
		\:\,\text{and}\:\, % & 
		A_2^{\rm reg} %&
		=-\frac{1}{128}
		\sum_{l=0}^\infty  (\nu z)^{-2\ep} \sum_{k=1}^3
		\,q_k \left(z^{2k-7}-1\right)\,.  
	}
	Let us recall
	that $z$ is defined in Equation \Eq{3.4.6} and $\sum_{k=1}^3
	\,q_k =(1 - 2 \ep)^2 (1 + 10 \ep)$.
	\textls[-25]{In $A_2^{\rm as}$, the summation over $l$ is explicit in the limit of $\ep\to0$.  Let us 
		use an integral representation, i.e.,}
	\eq{3.4.10a}{\sum_{l=0}^\infty (\nu^2+m^2)^{-\ep} = \frac{1}{\Gamma(\ep)}\int_0^{\infty}dt\, t^{\ep-1}e^{-m^2 t}\sum_{l=0}^{\infty}e^{-\nu^2 t}
		.  
	}
	The sum on the righ-hand side can 
	be expressed in terms of Jacobi theta functions $\theta_2(0|\tau)$ and $\theta_4(0|\tau)$, which obey the following relation~\cite{Whittaker_Watson_2021}:
	$$(-i\tau)^{1/2}\theta_2(0|\tau)=\theta_4(0|-1/\tau).$$
	Using this property we derive
	\eq{3.4.10b}{&\sum_{l=0}^\infty (\nu^2+m^2)^{-\ep} \\&= \frac{\sqrt{\pi}}{2\Gamma(\ep)}
		\left[%\{
		m^{1-2\ep}\Gamma\left(\ep-\frac12\right)+4 \left(\frac{m}{\pi}\right)^{\frac12-\ep}
		\sum_{l=1}^{\infty}(-1)^l l^{-\ep-\frac12}K_{\frac12-\ep}(2 m \pi l)
		\right]%\} 
		\nn
		,%.  
	}	
	where $K_x$ denotes the modified Bessel function of the second kind of order $x$. 
	The
	~representation \Eq{3.4.10b} 
	allows for the limit of $\ep\to0$, and one obtains
	\eq{3.4.11} {
		A_2^{\rm as} &= \mathcal{O} (\epsilon).
	}
	This way, there is no pole contribution and also no regular contribution from $A_2^{\rm as}$.
	
	In the regular part of $A_2^{\rm reg}$
	~\Eq{3.4.10},  we take $\ep=0$ directly, since the sum over $l$ converges. In this limit, let us 
	mention that
	\eq{3.4.12}{ q_1%&
		=-15\,%~~
		,\quad  %& 
		q_2&=24\,%~~, 
		,\quad\text{and}\quad %&
		q_3%&
		=-8\,%~~
		.
	} 
	With these 
	~numbers \Eq{3.4.12}, $A_2^{\rm reg}$~\Eq{3.4.10} 
	matches $A_2$ in Equation (A11) in~Ref. \cite{bord97-56-4896}.
	The sum representation \Eq{3.4.10} is convenient for numerical evaluation as long as $m$ is not 
	exceptionally
	large. The~asymptotic expansion for large enough $m$ values can be obtained with the \mbox{Euler--Maclaurin formula, i.e.,}
	\eq{A1}{ \sum_{l=0}^\infty  g(l) &= \int_0^\infty dl\, g(l)+\frac{g(0)}{2}-\sum_{j=1}^\infty\frac{B_{2j}}{(2j)!}g^{(2j-1)}(0),
	} 
	where $B_{2k}$ represents the Bernoulli numbers.  Taking 
	in Equation \Eq{3.4.10} 
	the sum over $k$ 
	for 
	\begin{equation}
		g(l)=\sum_{k=1}^3
		\,q_k \left(z^{2k-7}-1\right),
	\end{equation}
	with $z$ defined in Equation \Eq{3.4.6} and $q_k$ values 
	in Equation \Eq{3.4.12}, one arrives 
	at
	\eq{3.4.13}{ A_2^{\rm reg} &%\sim 
		\approx \frac{1}{384m}+\mathcal{O}%O
		\left(\frac{1}{m^2}\right).} 
	The term linear in $m$ and the constant terms cancel under the sum over $k$, and  one is left with the 
	next~order
	.
	The expression for $m=0$ follows like in the previous cases of $A_k$ and turns out to be
	\eq{3.4.14}{ A_2^{m=0} &=
		-\frac{1}{16} \left(2^{2 s+1}-1\right) s^2 (5 s+3) \zeta (2 s+1).
	}
	\subsection{$A_{3}$}
	\label{T35}
	Let us 
	start from Equation \Eq{3.8} with $k=3$ and note that 
	\eq{3.5.1}{ D_3(t) &= \frac{-5525 t^9+9945 t^7-4779 t^5+375 t^3}{5760}.
	}
	The result of the integration over $z$ in Equation \Eq{3.8} can be written in the form of
	\eq{3.5.2}{ A_3 &= \sum_{l=0}^\infty \nu(\nu^2+m^2)^{-4-\ep}P(\nu),
	}
	where
	\eq{3.5.3}{ P(\nu) &=
		\frac{c_3}{30240}\left(7875 m^6-189 (354\ep+229) m^4 \nu ^2 +63 (19 + 528 \ep + 884 \ep^2)  m^2 \nu ^4\right.
		\\\nn &\left.+(-687 + 2930 \ep + 2652 \ep^2 - 8840 \ep^3) \nu ^6\right)
	}
	is a polynomial in $\nu$, and~the $c_3$ factor is a function of $\ep$, i.e.,
	\eq{3.5.3a}{c_3 & =\frac{\cos (\ep \pi)}{\pi^{3/2}}
		\Gamma \left(\frac{3}{2}-\ep \right) \Gamma(\ep+1), \quad c_3|_{\ep\to0}=\frac{1}{2 \pi}.
	}
	One can rearrange $P(\nu)$ as follows:
	\eq{3.5.4}{  P(\nu) &= c_3\sum_{k=1}^4 b_k \, \tilde z^{k-1}, 
		\quad \tilde z=1+\frac{m^2}{\nu^2},
	}
	where
	\eq{3.5.4a}{
			b_1%&
			=-\frac{221}{756}(1+\ep)(2+\ep)(3+\ep), \; b_2=\frac{221}{120} (1+\ep)(2+\ep), \; b_3=-\frac{177}{80} (1+\ep), \; 
			\text{and}\; b_4=\frac{25}{96}.
		}
	Thus, one arrives  
	at
	\eq{3.5.5}{ A_3 &=c_3 \sum_{l=0}^\infty m^{-2\ep}(\tilde z-1)^\ep \,\tilde z^{-\ep}
		\sum_{k=1}^4  \frac{b_k}{\nu}\,\tilde z^{k-5}.
	}
	Here, it is convenient to split $ A_3$ into
	\eq{3.5.6}{ A_3 &= A_3^{\rm as}+A_3^{\rm sub},
	}
	where
	\eq{3.5.7}{  A_3^{\rm as} &=  c_3\sum_{l=0}^\infty m^{-2\ep}(\tilde z-1)^\ep \,\tilde z^{-\ep}
		\sum_{k=1}^4  \frac{ b_k}{\nu} ,
		\\\nn
		A_3^{\rm sub} &= c_3\sum_{l=0}^\infty  m^{-2\ep}(\tilde z-1)^\ep \, \tilde z^{-\ep}
		\sum_{k=1}^4  \frac{b_k}{\nu} \,\left(%[
		\tilde z^{k-5}-1\right)%]
		.
	}
	In $ A_3^{\rm as}$, the sum over $l$ is explicit, and for $\ep\to0$, one gets 
	\eq{3.5.8}{ A_3^{\rm as}&=-\frac{229}{40320 \pi  {\ep}}+\frac{2152-687 \gamma -2061 \log %(
			2%)
		}{60480 \pi }+\mathcal{O}%O
		\left(\ep^1\right).
	}
	In $A_3^{\rm sub}$, one may directly set 
	$\ep=0$, since the sums converge, and arrives at
	\eq{3.5.9}{ 	A_3^{\rm sub}  &=\sum_{l=0}^\infty \frac{1}{\pi\nu}%\left
		\Bigg[
		\frac{25}{192} \left(\frac{1}{\tilde z}-1\right)
		-\frac{177}{160} \left(\frac{1}{\tilde z^2}-1\right)
		+\frac{221}{120} \left(\frac{1}{\tilde z^3}-1\right)	
		-\frac{221 }{252  }\left(\frac{1}{\tilde z^4}-1\right)%\right
		\Bigg]
		.
	}
	Combined, Equations \Eq{3.5.8} and \Eq{3.5.9} match $A_3$ in Equation~(A11) in~Ref. \cite{bord97-56-4896}.
	
	The 
	formulas obtained just above 
	are sufficient for the pole part and for numerical evaluation of the regular part. 
	However, as~it turns out, one can do more and take the sums explicitly in terms of polygamma functions. The~result reads: 
	\eq{3.5.10}{ A_3^{\rm sub}&=
		\frac{1}{120960 \pi }\Bigg\{%[
		687 \psi ^{(0)}\left(\frac{1}{2}-i m\right)+687 \psi ^{(0)}\left(i m+\frac{1}{2}\right)
		\\\nn&	+i m \left[%(
		1023 \psi
		^{(1)}\left(\frac{1}{2}-i m\right)-1023 \psi ^{(1)}\left(i m
		+\frac{1}{2}\right)
		\right.
		\\\nn&\left.		+221 m \left(12 i \psi
		^{(2)}\left(\frac{1}{2}-i m\right)
		+12 i \psi ^{(2)}\left(i m+\frac{1}{2}\right)\right.\right.
		\\\nn&	\left.\left.	+5 m \psi^{(3)}\left(\frac{1}{2}-i m\right)
		-5 m \psi ^{(3)}\left(i m+\frac{1}{2}\right)\right)\right]%)
		+1374 (\gamma +\log 	%(
		4)%) 
		\Bigg\}%]
		.
	}
	Together with the regular part from $A_3^{\rm as}$ \Eq{3.5.8}, for the regular part  one finally gets 
	\eq{3.5.11}{ A_3^{\rm reg}&=
		\frac{1}{120960 \pi }\Bigg[
		1105 i m^3 \psi ^{(3)}\left(\frac{1}{2}-i m\right)-1105 i m^3 \psi ^{(3)}\left(i m+\frac{1}{2}\right)
		\\\nn&	-2652
		m^2 \psi ^{(2)}\left(\frac{1}{2}-i m\right)
		-2652 m^2 \psi ^{(2)}\left(i m+\frac{1}{2}\right)+1023 i m \psi
		^{(1)}\left(\frac{1}{2}-i m\right)
		\\\nn&	-1023 i m \psi ^{(1)}\left(i m+\frac{1}{2}\right)+687 \psi
		^{(0)}\left(\frac{1}{2}-i m\right)
		+687 \psi ^{(0)}\left(i m+\frac{1}{2}\right)
		\\\nn&	+4304-1374 \log %(
		2%)
		\Bigg]  
		\,.
	}
	Here, it is straightforward now to take the limit of $m\to\infty$, i.e.,
	\eq{3.5.12}{ A^{\mathrm inf}_3 &= -\frac{229 \left(\log %\left(
			\frac{2}{m}%\right)
			-1\right)}{20160 \pi }+\mathcal{O}%O
		\left(\left(\frac{1}{m}\right)^1\right).
	}
	For $m=0$, we proceed  as above in this Section 
	and receive
	\eq{3.5.13}{ A_3^{m=0} &=
		-\frac{1}{16} \left(2^{2 s+1}-1\right) s^2 (5 s+3) \zeta (2 s+1).
	}
	\section{Results for the Zeta~Function}\label{T4}
\begin{figure}[H]
		\includegraphics[width=13.2 cm]{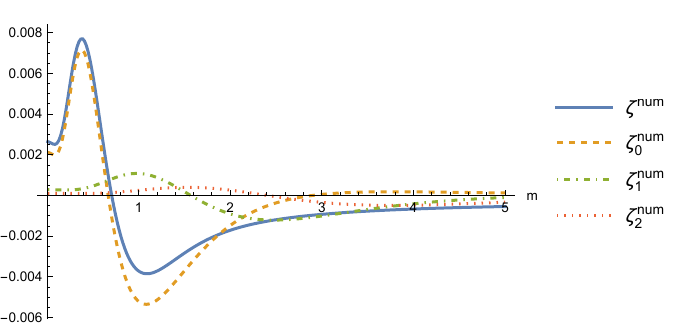}
		\caption{The $\zeta^{\rm num}$ \Eq{4.1} and $\zeta^{\rm num}_l$ \Eq{4.2} functions as functions of $m$.
		}		\label{fig:1}	
	\end{figure}

\begin{figure}[H]
		\includegraphics[width=13.2 cm]{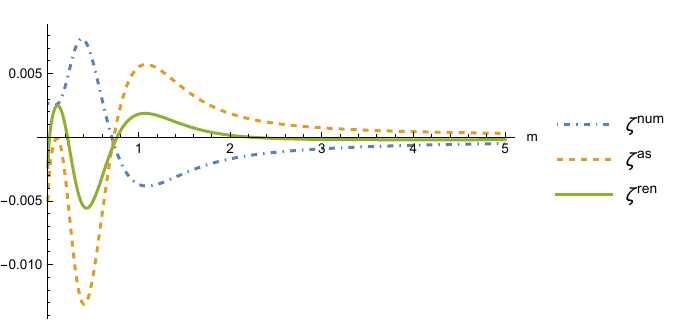}
		\caption{The $\zeta^{\rm ren}$ \Eq{4.3}, $\zeta^{\rm num}$, 
			and~$\zeta^{\rm as}$
			functions as~functions of $m$.
		}		\label{fig:2}	
	\end{figure}

	In this section, we summarize the results obtained above and present the basic plots.
	We start with the numerical part. For~the zeta function, i.e., Equation~\Eq{2.25}, it takes the form of
	\eq{4.1}{ \zeta^{\rm num}=2\sum_{l=0}^\infty \nu \,\zeta^{\rm num}_l,
	}
	where
	\eq{4.2}{  \zeta^{\rm num}_l &= -\frac{1}{\pi}
		\int_{m/\nu}^\infty dz
		\sqrt{(\nu z)^2-m^2} \frac{\pa}{\pa z}
		\left(	\ln \phi-\ln\phi^{\rm as}\right)
	}
	represents the contributions from the individual orbital modes. These are all functions of $m$, as shown in Figure~\ref{fig:1}, which reproduces 
	Figure~9.1 in~ Ref. \cite{BKMM}. For~fixed $m$ values, the~orbital contributions decrease in $l$, $ \zeta^{\rm num}_l\propto 
	l^{-3}$ 
	for $l\to\infty$, making the sum in Equation \Eq{4.1} convergent.
	For the asymptotic part, i.e., $\zeta^{\rm as}$ \Eq{3.7}, we have the individual contributions of $A_k=A_k^{\rm reg}$
	, 
	$k=-1,\dots,3$,
	~from Section \ref{T3}.
	Their
	sum, i.e.,
	\eq{4.3}{ \zeta^{\rm ren} &= \zeta^{\rm num}+\zeta^{\rm an},
	}
	is shown in Figure~\ref{fig:2}. There is a difference relative to Figure~9.2 in~Ref. \cite{BKMM}. The~analytic part there is larger, probably because of the mentioned 
	in Section \ref{T33} 
	sign error in $A_1(s)$ (Equation~(A11) in~Ref. \cite{bord97-56-4896}). As~a result, the~renormalized energy is for 
	large enough 
	$m$ values above and not below zero, as~also suggested by Equation~(9.2) in~Ref. \cite{BKMM}. Also, with~the correct numbers, the~renormalized zeta function at $m=0$ is still negative. At~$m=0$, we have $\zeta^{\rm 
		ren}=-0.0024$ with
	$\zeta^{\rm num}=0.0026$ and $\zeta^{\rm an}=-0.0050$. For~large $m$ values---$m=5$ in the 
	example here---we have
	$\zeta^{\rm ren}=-0.000199$, $\zeta^{\rm num}=-0.00050$, and~$\zeta^{\rm an}=0.00030$.
	\section{Conclusions}\label{T5}
	In
	this paper, we have presented 
	a thorough reconsideration of the vacuum energy of a massive scalar field obeying Dirichlet boundary conditions inside a spherical cavity, 
	as was originally studied in~Ref. \cite{bord97-56-4896}. We have repeated the complete calculation, corrected a sign error in one of the 
	analytic contributions appearing in the Appendix of Ref. \cite{bord97-56-4896}
	, and~expressed several  
	numerical sums from Ref. \cite{bord97-56-4896} in closed form using polygamma~functions.
	
	The correction has  significant
	consequences. It restores the physically expected behavior whereby the renormalized vacuum energy approaches zero in the large-mass limit, 
	in~agreement with the requirement that an infinitely heavy field not produce quantum fluctuations. For~
	relatively large $m$ values, it takes negative values, which resolves the recently observed discrepancy between Equation~(9.2) and 
	Figure~9.2 in~Ref. \cite{BKMM}.
	
\begin{figure}[H]
		\includegraphics[width=13.2 cm]{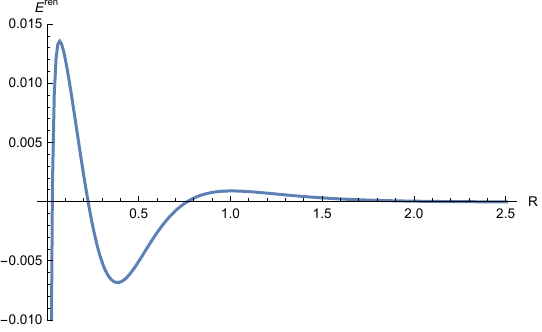}
		\caption{The
			renormalized vacuum energy ($E^{\rm ren}$
			~\Eq{2.16}) as~a function of the radius $R$ for particle mass $m=1$ (
			in $\hbar=c=1$ units).
		}		\label{fig:3}	
	\end{figure}

	Our numerical analysis reveals a rich structure in the renormalized vacuum energy as a function of the cavity radius 
	$R$
	(Figure \ref{fig:3}). Restoring the energy and mass
	dependences on $R$ ($E\to E/R$
	and  
	$m\to m R$, respectively) for a fixed mass ($m=1$ 
	$\hbar=c=1$ units), we found 
	that the energy changes sign four times, starting with a negative value for 
	quite small radii behaving 
	approximately as $
	-0.0012/R$. At~$m=0$ the renormalized zeta function is negative ($\zeta^{\rm ren}\approx -0.0024$), 
	while for relatively large $m$ values, both the numerical and analytic contributions become  considerably small, 
	with~the total value
	remaining~negative. 
	
	The renormalization scheme adopted here fixes the ambiguity by demanding that the renormalized energy vanish as $m\to\infty$. For~theories that can be formulated with a mass parameter, this procedure uniquely determines the massless limit by continuity. Remarkably, such a limit can be taken even in electrodynamics, despite the appearance of an additional polarization state when the photon acquires mass~\cite{bart84-311-336}. Finally, at~this point, 
	it is worth to
	mention that the counterterms defined by this scheme remain the same as in~Ref. \cite{bord97-56-4896}.
	The 
	results obtained
	are relevant not only for Casimir physics but also for the bag model in QCD and for heat kernel techniques on manifolds with 
	boundaries.
	 
	The~explicit expressions for the analytic parts of the zeta function obtained here 
	are believed to
	facilitate further numerical studies and possible extensions to other boundary conditions or~geometries.
	Future work may include the consideration of Robin boundary conditions, an~electromagnetic case with massive photons, or~applications to finite-temperature effects beyond those already~studied.

%\bibliographystyle{plain}
%\bibliography{D:/WORK/Literatur/bib/papers}

\end{document}